\documentclass[prc,twocolumn,showpacs]{revtex4}

\usepackage{graphicx}
\usepackage{latexsym} 

\newcommand{\be}{\begin{equation}}
\newcommand{\ee}{\end{equation}}
\newcommand{\ba}{\begin{eqnarray}}
\newcommand{\ea}{\end{eqnarray}}
\newcommand{\ban}{\begin{eqnarray*}}
\newcommand{\ean}{\end{eqnarray*}}

\begin{document}

\title{Deciphering Azimuthal Correlations \\ 
in Relativistic Heavy-Ion Collisions}

\author{Tomasz Cetner and Katarzyna Grebieszkow}

\affiliation{Faculty of Physics, Warsaw University of Technology,
ul. Koszykowa 75, 00-662 Warszawa, Poland}

\author{Stanis\l aw Mr\'owczy\'nski}

\affiliation{Institute of Physics, Jan Kochanowski University, 
ul. \'Swi\c etokrzyska 15, PL - 25-406 Kielce, Poland \\
and So\l tan Institute for Nuclear Studies,
ul. Ho\.za 69, PL - 00-681 Warsaw, Poland}

\date{January 18, 2010}

\begin{abstract}

We discuss various sources of azimuthal correlations in relativistic heavy-ion collisions.
The integral measure $\Phi$ is applied to quantify the correlations. We first consider separately 
the correlations caused by the elliptic flow, resonance decays, jets and transverse momentum 
conservation. An effect of randomly lost particles is also discussed. Using the PYTHIA and 
HIJING event generators we produce a sample of events which mimic experimental data.  
By means of kinematic cuts and particle's selection criteria, the data are analyzed to identify  
a dominant source of correlations. 

\end{abstract}

\vspace{0.3cm}


%25.75.-q Relativistic heavy-ion collisions
%25.75.Gz Particle correlations

%\vspace{0.5cm}

\maketitle

%%%%%%%%%%%%%%%%%%%%%%%%%%%%%%%%%%%%%%%%%%%%%%%%%%%%%%%%%%%%
\section{Introduction}
%%%%%%%%%%%%%%%%%%%%%%%%%%%%%%%%%%%%%%%%%%%%%%%%%%%%%%%%%%%%

Particles produced in relativistic heavy-ion collisions are correlated in azimuthal
angle due to various mechanisms. One mentions here extensively studied jets and
minijets resulting from (semi-)hard parton-parton scattering and collective flow
due to the cylindrically asymmetric pressure gradients, see the review articles
\cite{CasalderreySolana:2007zz} and \cite{Voloshin:2008dg}, respectively. More
exotic sources of correlations are also possible. As argued in \cite{Mrowczynski:2005gw},
the plasma instabilities, which occur at an early stage of collisions, can generate
the azimuthal fluctuations. Except the dynamically interesting mechanisms, there
are also rather trivial effects caused by decays of hadronic resonances or by
energy-momentum conservation.

There is a variety of methods designed to study fluctuations on event-by-event basis.
In particular, the so-called measure $\Phi$ proposed in \cite{Gazdzicki:1992ri}
was used to measure the transverse momentum \cite{Anticic:2003fd,Anticic:2008vb} 
and electric charge fluctuations \cite{Alt:2004ir}. The measure proved to be very sensitive 
to dynamical correlations and it was suggested to apply it to study azimuthal ones
\cite{Mrowczynski:1999vi}. Such an analysis is underway using experimental data accumulated 
by the NA49 and NA61 Collaborations and some preliminary results are already published
\cite{Cetner:2010vz}.  The aim of this paper is to present model simulations to be used in 
interpretation of the experimental data. The fact that the measure $\Phi$ is sensitive to 
correlations of various origin is advantage and disadvantage at the same time, as it is difficult 
to disentangle different contributions. Therefore, we model the azimuthal correlations 
driven by several processes and we look how the correlations show up when quantified 
by the measure $\Phi$. We first consider separately in terms of toy models, the elliptic flow, 
resonance decays, jets and transverse momentum conservation. An effect of randomly lost 
particles is also examined. Then, we analyze the data provided by the PYTHIA and HIJING
event generators showing how to identify the main sources of correlations by applying 
kinematic cuts and  particle's selection criteria.

A magnitude of $\Phi$ of azimuthal correlations measured in Pb-Pb collisions at 158$A$ GeV 
is of order of 0.01 radian \cite{Cetner:2010vz}. Since the effects, which are theoretically studied 
here, often exceed this order, a realistic model should include {\em all} the effects, and 
consequently, the model  has to be rather complex. For this reason we do not attempt to 
compare our model calculations to the preliminary data \cite{Cetner:2010vz}. 
There is also an additional reason. The measurement \cite{Cetner:2010vz} has been 
performed in an acceptance window which is not only limited in particle's rapidity and 
transverse momentum but it is also nonuniform in azimuthal angle. The acceptance, which 
is characteristic for the NA49 detector, has to be properly included in model calculations 
for a quantitative comparison with the experimental data. It is beyond the scope of this study.
Our aim here is to understand how the collective flow, resonance decays, jets and transverse 
momentum conservation contribute to azimuthal correlations and how the contributions 
can be disentangled.

%%%%%%%%%%%%%%%%%%%%%%%%%%%%%%%%%%%%%%%%%%%%%%%%%%%%%%%%%%%
\section{Measure $\Phi$}
%%%%%%%%%%%%%%%%%%%%%%%%%%%%%%%%%%%%%%%%%%%%%%%%%%%%%%%%%%%%

Let us first introduce the correlation measure $\Phi$. One defines the variable
$z\buildrel \rm def \over = x - \overline{x}$, where $x$ is a single particle's
characteristics such as the particle transverse momentum, electric charge or azimuthal 
angle. The overline denotes averaging over a single particle inclusive distribution.
In the subsequent sections, $x$ will be identified with the particle azimuthal angle
$\phi$ and the fluctuation measure will be denoted as $\Phi_\phi$. The event 
variable $Z$, which is a multiparticle analog of $z$, is defined as 
$Z \buildrel \rm def \over = \sum_{i=1}^{N}(x_i - \overline{x})$, where the summation
runs over particles from a given event. By construction, $\langle Z \rangle = 0$,
where $\langle ... \rangle$ represents averaging over events (collisions). The
measure $\Phi$ is finally defined as
\be
\label{phi-def}
\Phi \buildrel \rm def \over =
\sqrt{\langle Z^2 \rangle \over \langle N \rangle} -
\sqrt{\overline{z^2}} \;.
\ee
It is evident that $\Phi = 0$, when no inter-particle correlations are present.
The measure also possesses a less trivial property - it is {\it independent} of
the distribution of the number of particle sources if the sources are identical
and independent from each other. Thus, the measure $\Phi$ is `blind' to the
impact parameter variation as long as the `physics' does not change with the
collision centrality. In particular, $\Phi$ is independent of the impact parameter
if the nucleus-nucleus collision is a simple superposition of nucleon-nucleon
interactions. In the following sections we discuss how various mechanisms 
responsible for azimuthal correlations contribute to $\Phi_\phi$.  Then, using the
event generators we show how the dominant contributions can be identified.

\begin{figure*}[t]
\centering
\vspace{-1cm}
\includegraphics*[width=0.49\textwidth]
{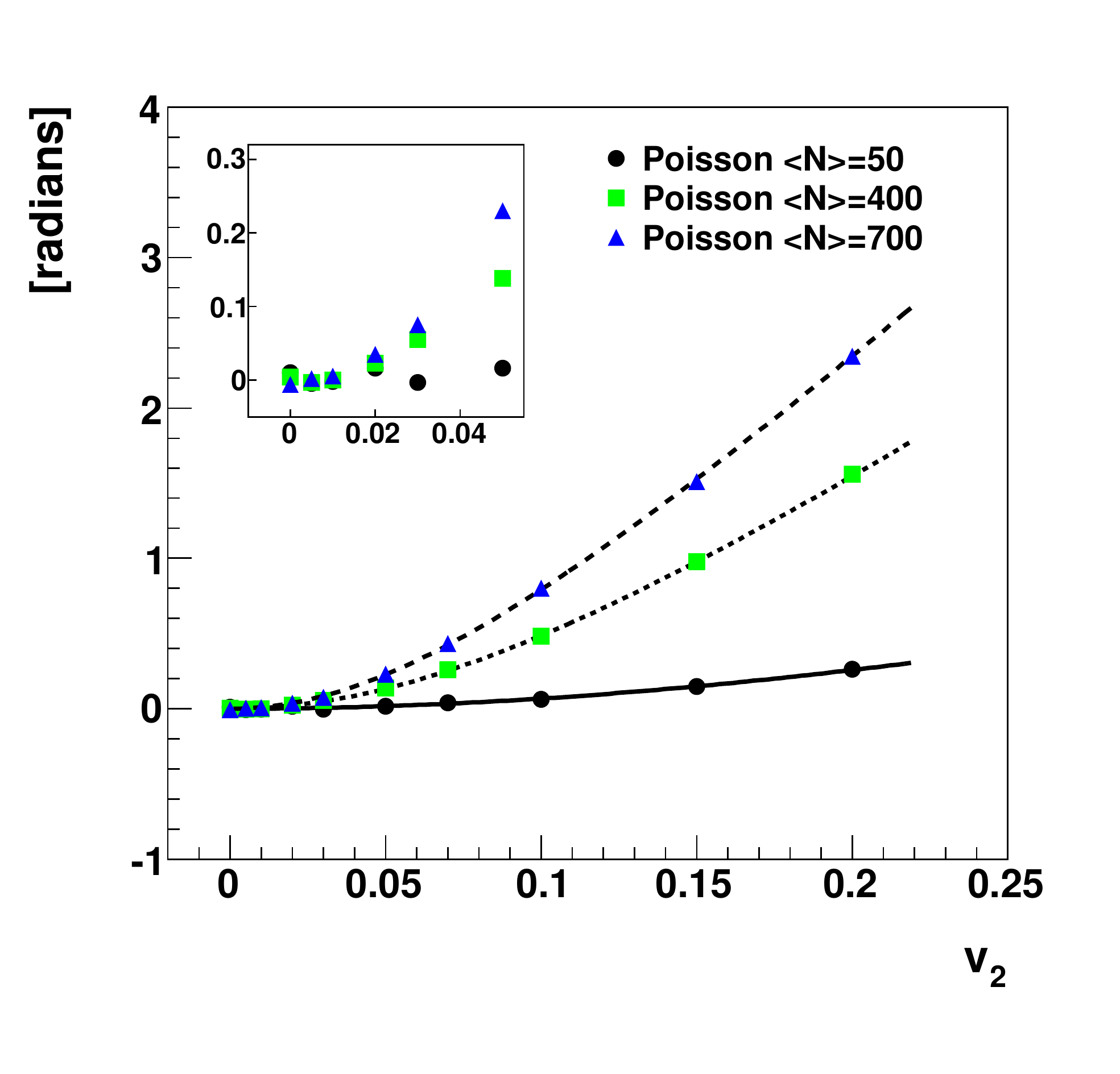}
\includegraphics*[width=0.49\textwidth]
{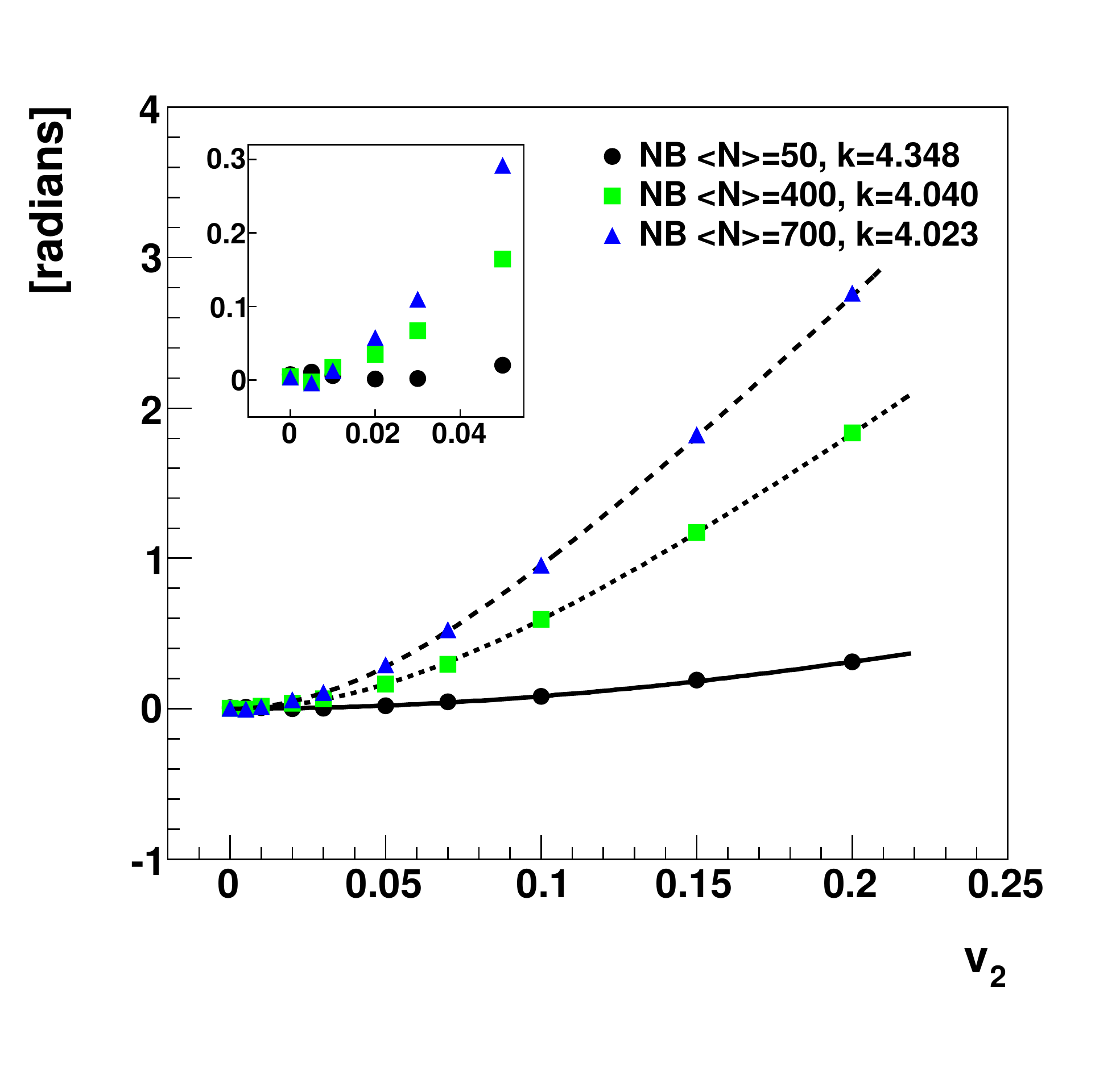}
\vspace{-1cm}
\caption{(Color online) $\Phi_\phi$ as a function of the second Fourier
coefficient $v_2$ for the Poisson (left panel) and NB (right
panel) multiplicity distributions. The lines represent
the analytical formula (\ref{phi-col-exact}). The insets show 
$\Phi_\phi$ for small values of $v_2$. }
\label{fig-v2only}
\end{figure*}

%%%%%%%%%%%%%%%%%%%%%%%%%%%%%%%%%%%%%%%%%%%%%%%%%%%%%%%%%%%
\section{Collective flow}
%%%%%%%%%%%%%%%%%%%%%%%%%%%%%%%%%%%%%%%%%%%%%%%%%%%%%%%%%%%%

Particles produced in relativistic heavy-ion collisions reveal a collective
behavior which is naturally described in terms of hydrodynamics where 
the collective flow is caused by pressure gradients. The collective flow as 
quantified by the measure $\Phi_\phi$ was studied in \cite{Mrowczynski:1999vi}.
Here we only recapitulate the results derived in \cite{Mrowczynski:1999vi}.

Since the inclusive azimuthal distribution is flat, 
$\overline \phi = \pi$ and $\overline{\phi^2} = {4 \over 3}\pi^2$ for 
$\phi \in [0,2\pi]$, and thus $\overline{z^2} = {1 \over 3}\pi^2$. A single 
particle azimuthal distribution in a given event is 
\be
\label{event-dis}
P(\phi) = {1 \over 2\pi} \;
\Big[1 + 2\sum_{n=1}^{\infty} v_n {\rm cos}\big(n(\phi- \phi_R)\big) \Big] ,
\ee
where $0 \le \phi \le 2\pi$; $\phi_R$ is the reaction plane angle and $v_n$ denotes 
an amplitude of the $n-$th Fourier harmonics. The $N-$particle distribution is 
assumed to be a product on $N$ distributions (\ref{event-dis}) multiplied by a 
multiplicity distribution. Consequently, the collective flow is the only source of azimuthal 
correlations in the system. Averaging over particles is performed by integrating 
over $\phi_i$ with $i=1,\;2,\; \dots N$ and averaging over events is achieved by 
integrating over $\phi_R$ and summing over $N$. The distribution of reaction plane 
angle is obviously flat. Thus, one finds the measure $\Phi$ of azimuthal correlations 
caused by the flow as
\be
\label{phi-col-exact}
\Phi_\phi = \sqrt{ {\pi^2 \over 3} +  \bigg({\langle N^2 \rangle
- \langle N \rangle \over \langle N \rangle }\bigg)\, S }
- {\pi \over \sqrt{3}} .
\ee
where $\langle N^m \rangle$ is the $m-$th moment of multiplicity
distribution and
\be
S \equiv 2 \Big\langle \sum_{n=1}^{\infty}
\Big({v_n \over n} \Big)^2 \Big\rangle .
\ee

We have first verified the effect of second Fourier coefficient $v_2$ on $\Phi_\phi$ 
by Monte Carlo simulations. For this purpose we have generated events of particle
multiplicity given by either Poisson or Negative Binomial (NB) distribution.
The latter is defined as
\be
P_N = \frac{\Gamma (N+k)}{ \Gamma (N+1) \Gamma (k)}
\frac{\langle N \rangle^N k^k}
{\big(\langle N \rangle +k\big)^{N+k}} ,
\ee
where $\Gamma (k)$ is the Gamma function, which for positive integer arguments
equals $\Gamma (k) = (k-1)!$; the parameter $k$ can be expressed through
the variance of the distribution
${\rm Var}(N) \equiv \langle N^2 \rangle - \langle N \rangle^2$
and the average value $\langle N \rangle$ as
\be
k = \frac{\langle N \rangle^2}
{{\rm Var}(N) - \langle N \rangle} .
\ee
The parameter $k$ is chosen in such a way in our all simulations that
$\sqrt{{\rm Var}(N)} = \langle N \rangle /2$. Then, the multiplicity
distribution approximately obeys the Wr\'oblewski's formula 
\cite{Wroblewski:1973tn} which is known to hold for proton-proton
interactions in a wide collision energy range.  The simulations are
performed for both the Negative Binomial and Poisson distributions 
as the former distribution is much broader than the latter one for 
$ \langle N \rangle \gg 1$.  We note here that the width of multiplicity 
distributions in relativistic heavy-ion collisions strongly depends on
centrality selection criteria.  Thus, it is important to see how the 
correlation signal changes with the width of multiplicity distribution.

\begin{figure}[b]
\centering
\vspace{-0.7cm}
\includegraphics*[width=0.49\textwidth]{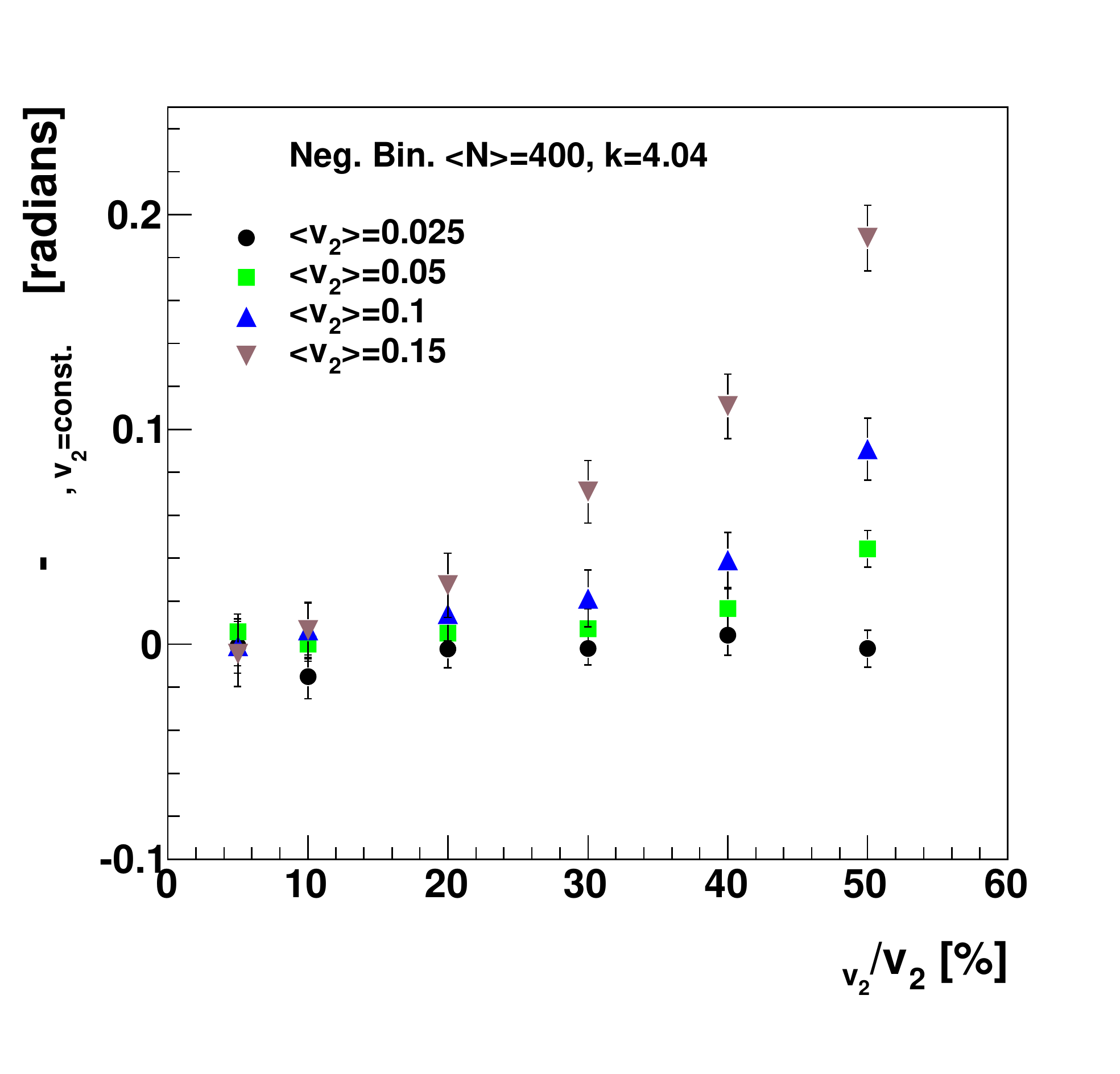}
\vspace{-1.2cm}
\caption {(Color online) The difference of $\Phi_\phi$ computed for the fluctuating and  
fixed $v_2$. }
\label{fig-fifi_v2_varN_rel}
\end{figure}

\begin{figure*}[t]
\centering
\vspace{-1cm}
\includegraphics*[width=0.49\textwidth]
{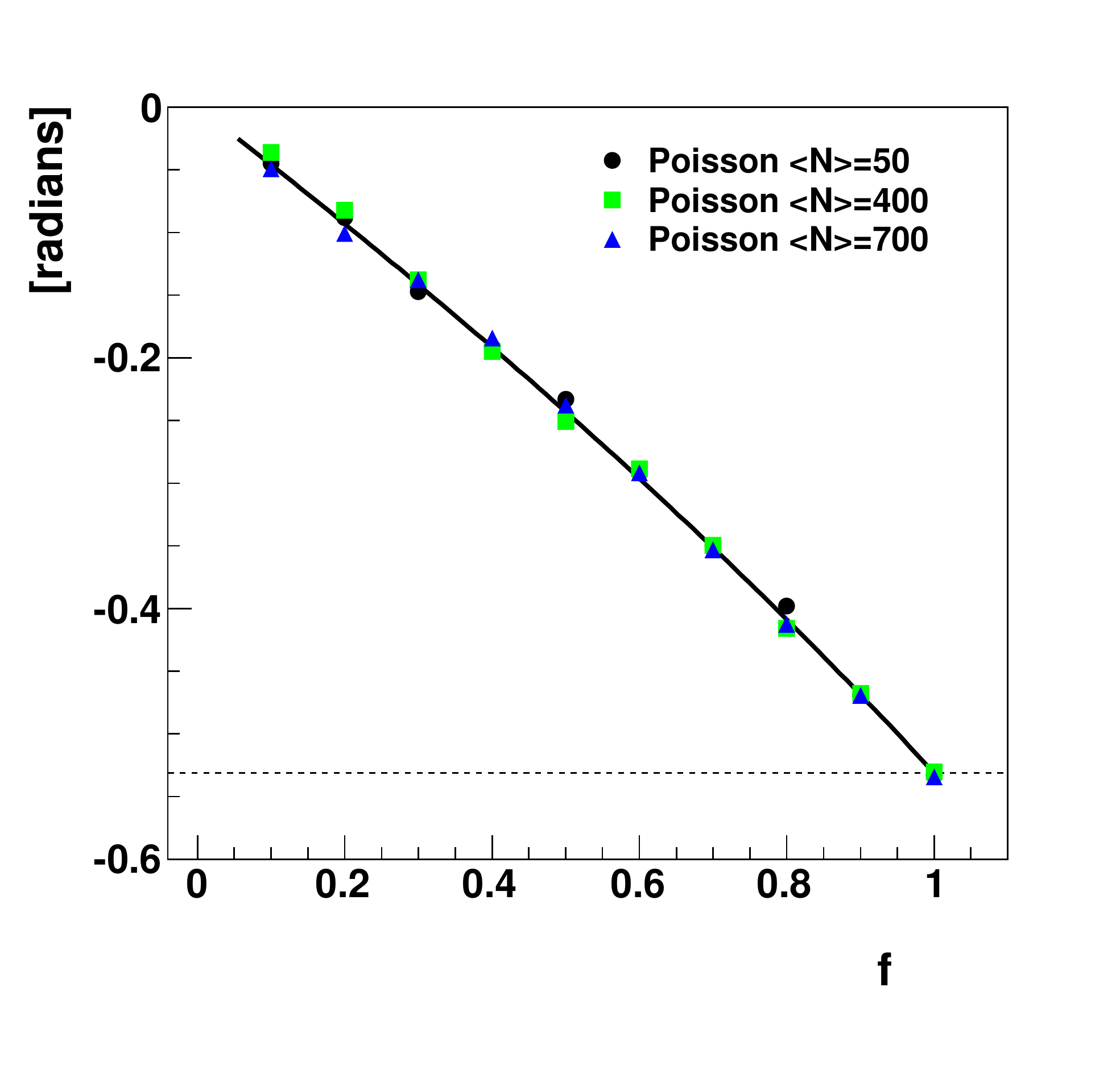}
\includegraphics*[width=0.49\textwidth]
{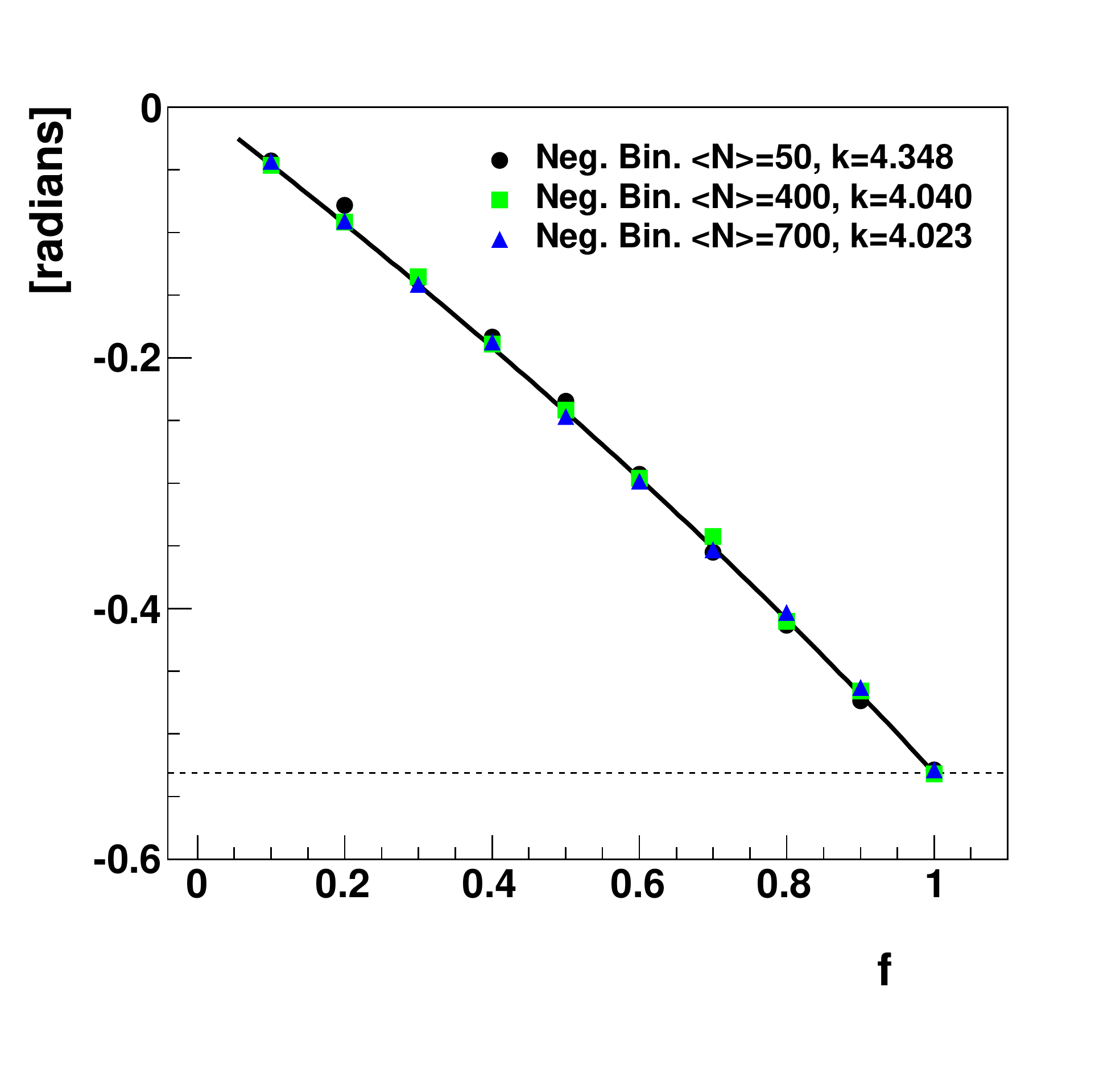}
\vspace{-1cm}
\caption{(Color online) $\Phi_{\phi}$ as a function of fraction of particles coming from 
the back-to-back resonance decays. The particle multiplicity is distributed 
according to the Poisson (left panel) or NB (right panel) distribution. The 
dashed and solid lines represent the analytical formulas (\ref{Phi-resonance-decays})
and (\ref{Phi-resonance-decays-fraction}), respectively.}
\label{fig-resfrac}
\end{figure*}

The azimuthal angle of each particle has been generated from the distribution
\be
\label{v2-dis}
P(\phi)= \frac{1}{2\pi} \Big(1 + 2 v_2 {\rm cos}\big(2(\phi - \phi_R) \big) \Big) ,
\ee
where $0 \le \phi \le 2\pi$; the reaction plane angle $\phi_R$ of a given event 
has been generated from the flat distribution. The results of our simulations are 
shown in Fig.~\ref{fig-v2only} for both the Poisson (left panel) and NB (right panel) 
multiplicity distributions. As seen, the analytical formula (\ref{phi-col-exact}) works 
perfectly well. 

There are large ($\sim 40\%$) event-by-event fluctuations of $v_2$ observed 
\cite{Voloshin:2008dg} at BNL RHIC. The $v_2$ fluctuations are 
dominated by the fluctuations of eccentricity of the overlap region of colliding 
nuclei, see e.g.  \cite{Broniowski:2007ft}. We have introduced the $v_2$ 
fluctuations in our simulations in the following way. For each event the value 
of $v_2$ has been generated from the Gaussian distribution of the dispersion 
$\sigma_{v_2}$. The fluctuations have been restricted to vary within 
$2 \sigma_{v_2}$ around the mean $\langle v_2 \rangle$ that is 
$\langle v_2 \rangle - 2 \sigma_{v_2} \le v_2 \le \langle v_2 \rangle + 2 \sigma_{v_2}$. 
Then, $v_2$ remains positive unless $\sigma_{v_2}/\langle v_2 \rangle$
exceeds 0.5.

In Fig.~\ref{fig-fifi_v2_varN_rel} we demonstrate the effect of flow fluctuations 
relative the effect of flow. Specifically, we show the difference of the correlation 
measures computed for the fluctuating $v_2$ and  fixed $v_2= \langle v_2 \rangle$.  
The particle multiplicity has been generated from the NB distribution with 
$\langle N \rangle =400$. As seen, the flow fluctuations of relative magnitude of 
$\sim 40\%$ noticeably increase the value of $\Phi_\phi$ if $\langle v_2 \rangle$ 
is not too small.

\begin{figure}[b]
\centering
\vspace{-0.6cm}
\includegraphics*[width=0.49\textwidth]{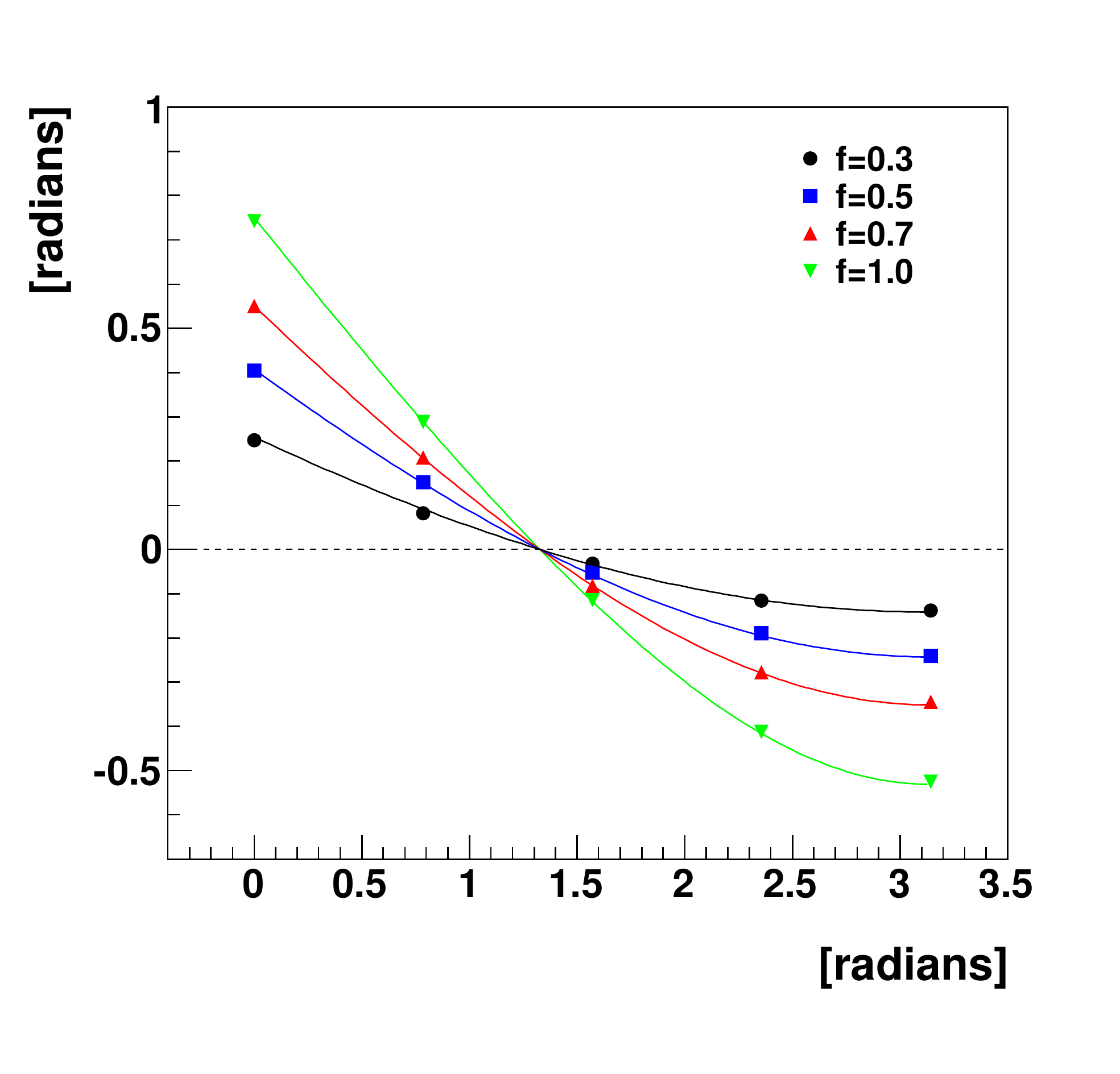}
\vspace{-1.0cm}
\caption{(Color online) $\Phi_\phi$ as a function of the correlation angle
$\Delta \phi$ for varying fraction $f$ of particles emitted in pairs.
The particle multiplicity is given by the NB distribution.}
\label{fig-fracres}
\end{figure}

%%%%%%%%%%%%%%%%%%%%%%%%%%%%%%%%%%%%%%%%%%%%%%%%%%%%%%%%%%%
\section{Resonance decays}
%%%%%%%%%%%%%%%%%%%%%%%%%%%%%%%%%%%%%%%%%%%%%%%%%%%%%%%%%%%%

\begin{figure*}[t]
\centering
\vspace{-1cm}
\includegraphics*[width=0.49\textwidth]
{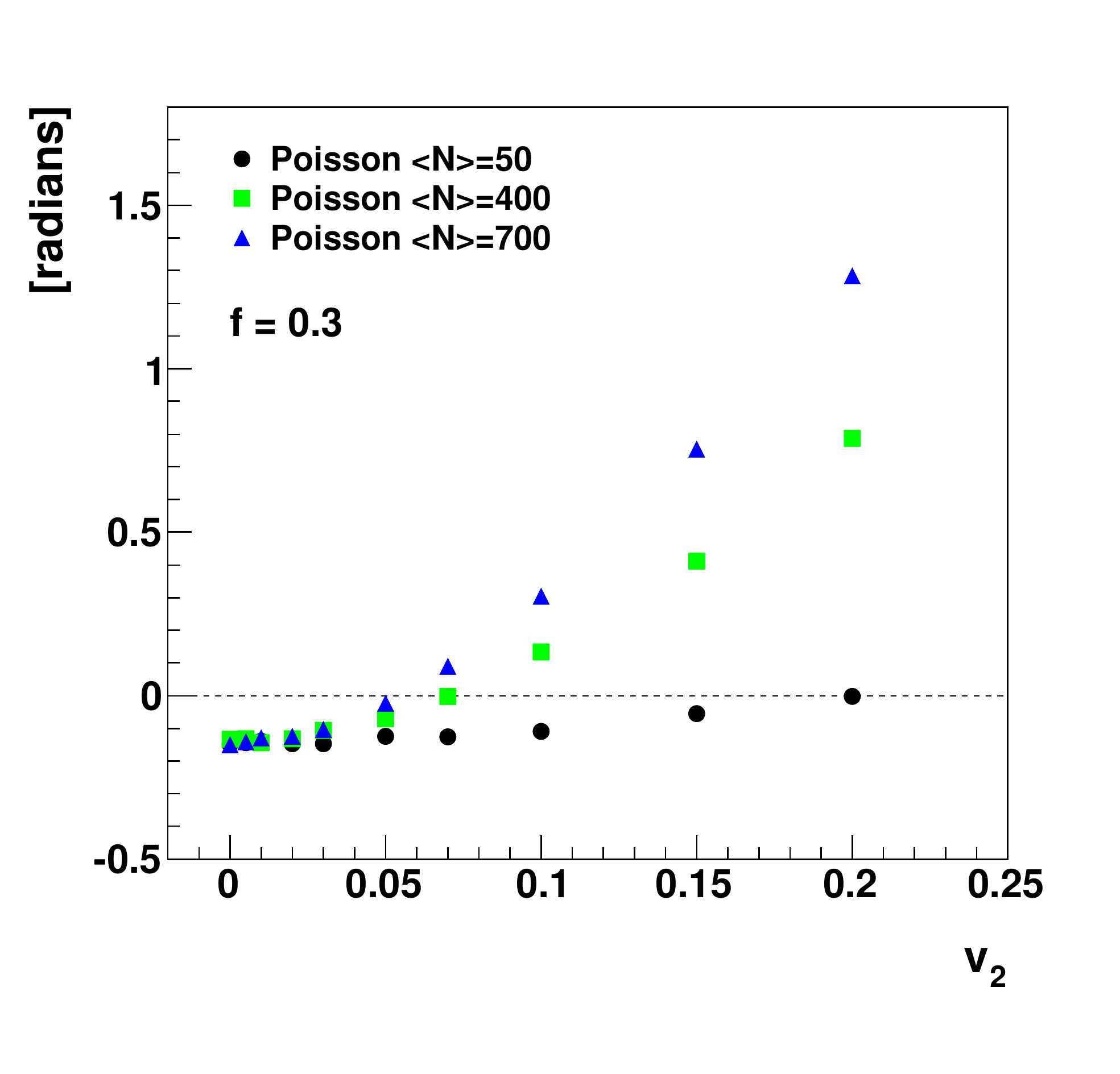}
\includegraphics*[width=0.49\textwidth]
{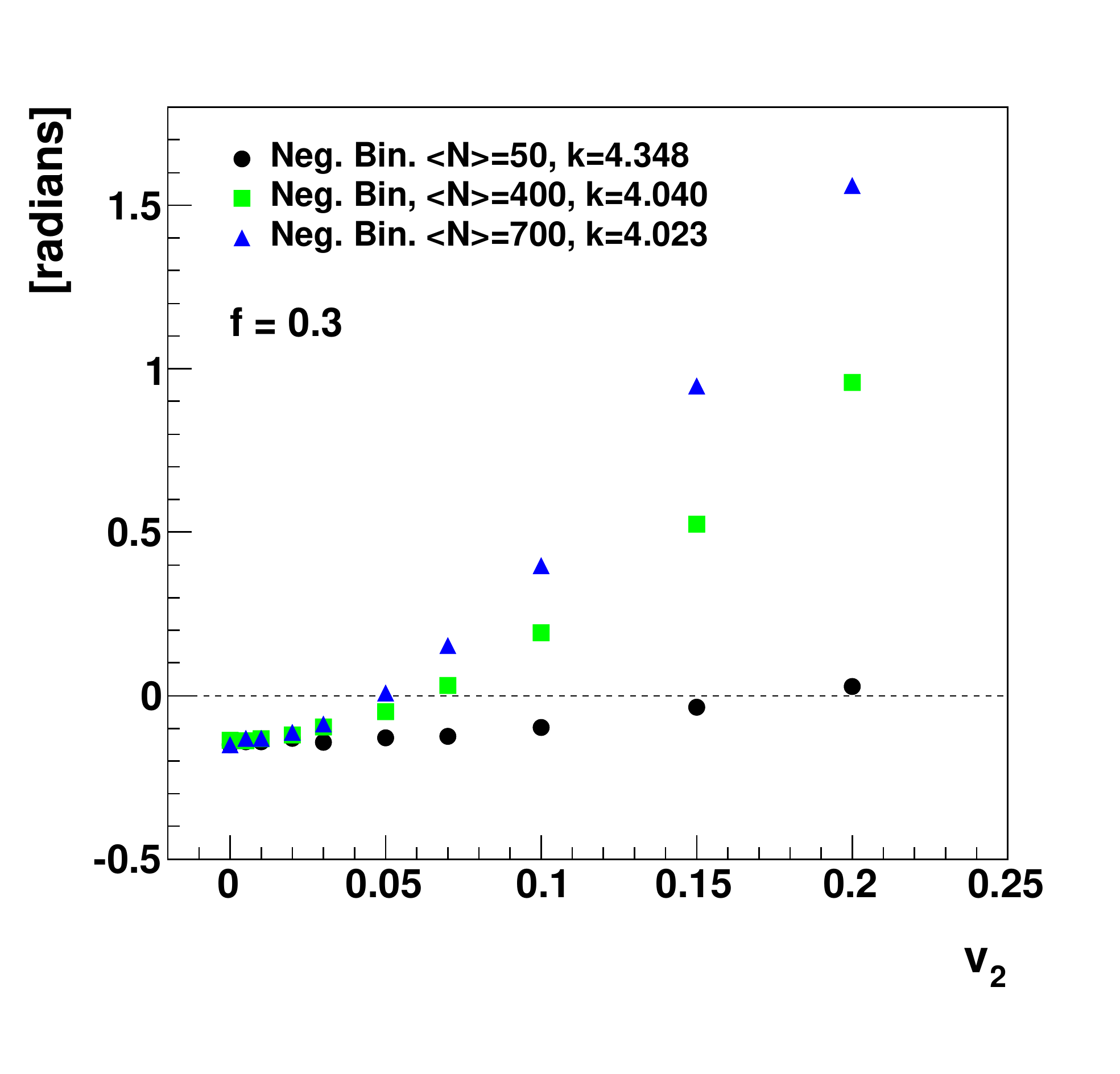}
\vspace{-1cm}
\caption{(Color online) $\Phi_\phi$ resulting from the combined effect of
elliptic flow and resonance decays. The particle multiplicity is
distributed according to the Poisson (left panel) or NB
(right panel) distribution.}
\label{fig-v2res}
\end{figure*}

Let us start the discussion of effects of resonance decays with the toy model 
where {\em all} produced particles come from heavy resonances which have 
vanishing transverse velocity and decay back to back into pairs of particles. 
The particle multiplicity is arbitrary but fixed even number. Then, as shown 
in Appendix, we have
\be
\label{Phi-resonance-decays}
\Phi_\phi = \frac{1 - \sqrt{2} }{\sqrt{6}}\, \pi \approx -0.531 .
\ee
When only a fraction $f$ of all produced particles comes from the
back-to-back decays of resonances while the remaining particles are 
produced independently from each other, the calculations presented
in Appendix lead to
\be
\label{Phi-resonance-decays-fraction}
\Phi_\phi = \frac{\sqrt{2-f} - \sqrt{2}}{\sqrt{6}} \pi \;.
\ee
As seen, for $f = 0$ the formula (\ref{Phi-resonance-decays-fraction})
gives, as expected, $\Phi=0$ and for $f = 1$ we get the value
(\ref{Phi-resonance-decays}).

We have checked the formula (\ref{Phi-resonance-decays-fraction})
by Monte Carlo simulations and then we have considered the model
where the particle multiplicity is not fixed but it is given by
either Poisson or NB distribution with the average multiplicity
equal 50, 400, or 700 particles.  For a given fraction $f$  of particles 
coming from the back-to-back decays of resonances,  the number
of particles coming from the decays in the event of multiplicity $N$ has 
been the even number which is the nearest to $f N$.  The measure $\Phi_\phi$
as a function of $f$ is shown in Fig.~\ref{fig-resfrac}. As seen, the formula 
(\ref{Phi-resonance-decays-fraction}) still works very well.

When a resonance, which is at rest, decays back to back, the
difference of azimuthal angles of the decay products is
$\Delta \phi = \pi$. When the resonance has a finite velocity, the
difference of azimuthal angles of the decay products is
smaller than $\pi$. When the resonance's kinetic energy is
much larger than the energy released in its decay, $\Delta \phi$
is zero. Therefore, we consider a model where a fraction of
particles comes from the resonance decays and the particles
are emitted in pairs with the difference of their azimuthal
angles $\Delta \phi$ varying from 0 to $\pi$. We first assume
that {\em all} particles are emitted in pairs and the relative 
azimuthal angle of two correlated particles equals $\Delta \phi$.
As shown in Appendix, we then have
\be
\label{Phi-resonance-decays-general}
\Phi_\phi = \sqrt{ \frac{2}{3} \pi^2 - \Delta \phi \, \pi + \frac{1}{2} (\Delta \phi)^2}
- \frac{\pi}{\sqrt{3}} .
\ee
As seen  that $\Phi_\phi$ changes its sign from positive to negative with growing 
$\Delta \phi$; $\Phi_\phi$ vanishes when 
\be
\label{zero-Phi}
 \Delta \phi = \pi \Big(1 - \frac{1}{\sqrt{3}} \Big)  \approx 1.328
\ee
and for $\Delta \phi = \pi$ we deal with the model described by the formula 
(\ref{Phi-resonance-decays}). Further on, we have considered a model where 
only a fraction $f$ of particles is emitted in correlated pairs. Then, as explained 
in Appendix, Eq.~(\ref{Phi-resonance-decays-general}) gets the form
\be
\label{Phi-resonance-decays-general-f}
\Phi_\phi = \sqrt{ \frac{\pi^2}{3} + f 
\Big( \frac{\pi^2}{3}  - \Delta \phi \, \pi + \frac{1}{2} (\Delta \phi)^2 \Big)}
- \frac{\pi}{\sqrt{3}} .
\ee
As previously, $\Phi_\phi$ changes its sign and $\Phi_\phi = 0$ for $ \Delta \phi$ given 
by Eq.~(\ref{zero-Phi}).

\begin{figure}[b]
\centering
\vspace{-1.0cm}
\includegraphics*[width=0.49\textwidth]{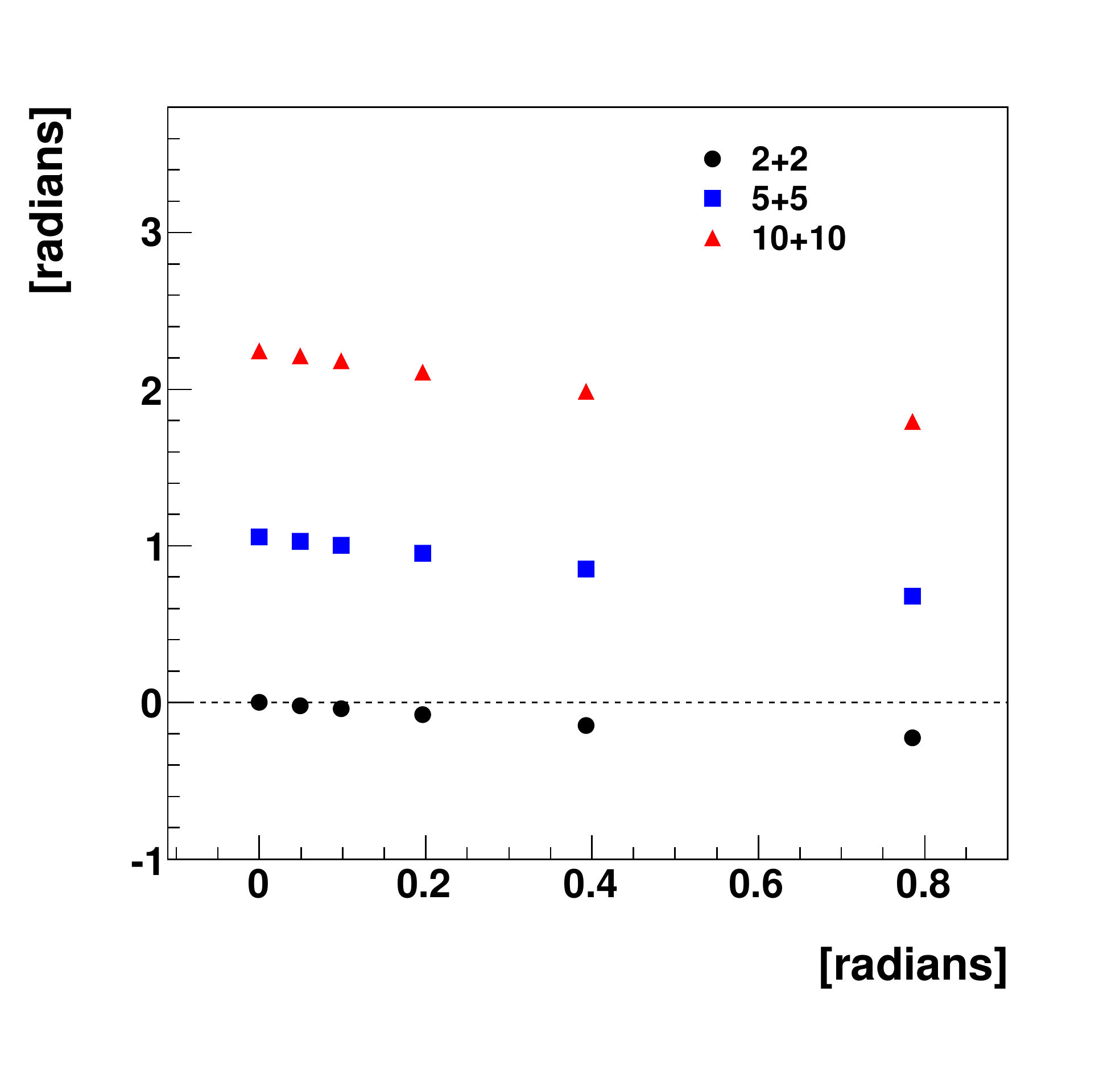}
\vspace{-1.0cm}
\caption {(Color online) $\Phi_\phi$ as a function of the jet opening
angle $\sigma \phi$ for several numbers of particles
in a jet. }
\label{fig-fifi_jet}
\end{figure}

In Fig.~\ref{fig-fracres}  we compare the formula (\ref{Phi-resonance-decays-general-f}) 
with the results of Monte Carlo simulation of $\Phi_\phi$ as a function of $\Delta \phi$. 
The fraction of particles emitted in pairs equals 0.3, 0.5, 0.7 or 1.0. The remaining 
particles, which are not emitted in pairs, carry no correlations. The particle multiplicity is
generated from the NB distribution with $\sqrt{{\rm Var}(N)} = \langle N \rangle /2 = 50$.
As seen, the formula (\ref{Phi-resonance-decays-general-f}) works perfectly well. 

\begin{figure*}[t]
\vspace{-1.0cm}
\begin{minipage}{8.5cm}
\centering
\vspace{-0.3cm}
\includegraphics*[width=8.5cm]{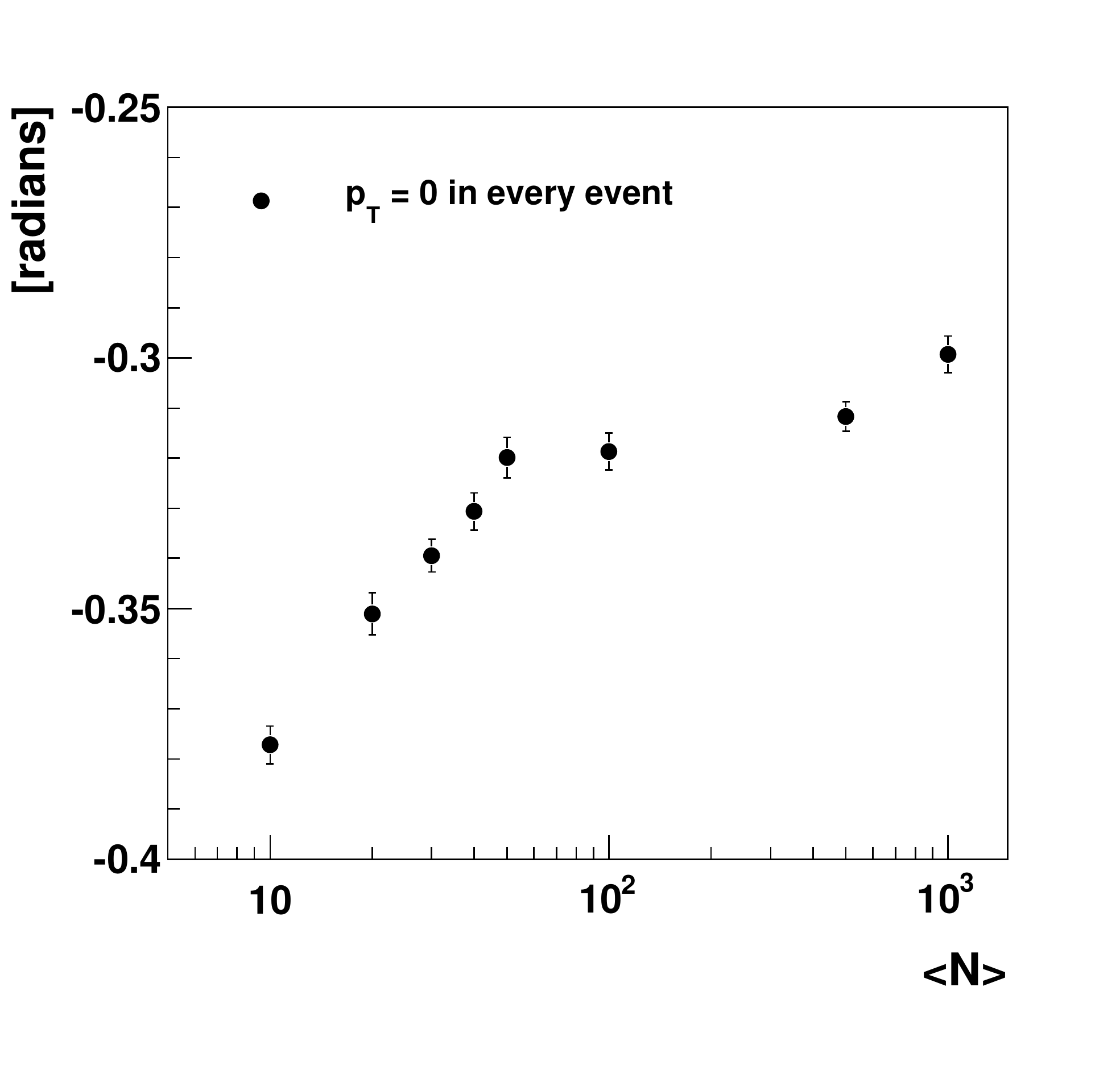}
\vspace{-1.3cm}
\caption{$\Phi_\phi$ as a function of average multiplicity
in events where total transverse momentum exactly vanishes.}
\label{fig-zach}
\end{minipage}\hspace{5mm}
\begin{minipage}{8.5cm}
\centering
\includegraphics*[width=8.5cm]
{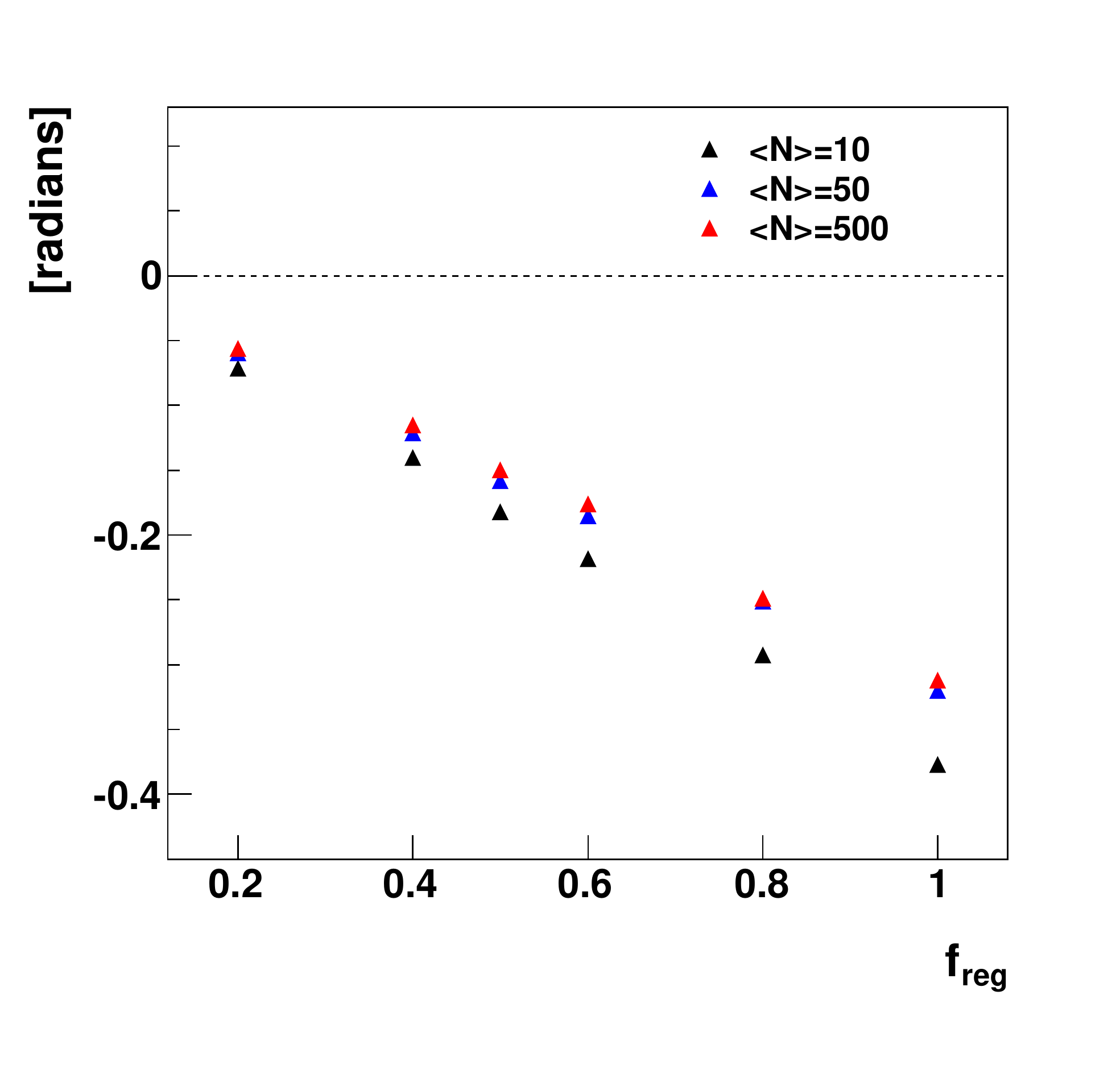}
\vspace{-1.3cm}
\caption {(Color online) $\Phi_\phi$ as a function of fraction of registered particles 
$f_{\rm reg}$.  The correlation results from the transverse momentum conservation.}
\label{zach_acc}
\end{minipage}
\end{figure*}

We next discuss the combined effect of resonance decays and elliptic flow.
The multiplicity of events is generated from the Poisson or NB binomial
distribution. For each event 30\% of particles is assumed to originate
from heavy resonances which decay back to back into pairs of particles.
Neither resonances nor their decay products experience any flow, but the
remaining 70\% of particles manifest the collective elliptic flow according
to Eq.~(\ref{v2-dis}). The results of the simulation are shown in
Fig.~\ref{fig-v2res}. When the particle multiplicity or $v_2$ are sufficiently
small, the effects of resonance decays dominates and $\Phi_\phi$ is negative.
It becomes positive when the effect elliptic flow takes over.

%%%%%%%%%%%%%%%%%%%%%%%%%%%%%%%%%%%%%%%%%%%%%%%%%%%%%%%%%%%
\section{`Dijets'}
%%%%%%%%%%%%%%%%%%%%%%%%%%%%%%%%%%%%%%%%%%%%%%%%%%%%%%%%%%%

We call a dijet the two groups (jets) of particles flying in exactly opposite
directions. Particles from each jet are distributed within a cone of the azimutal
angle $\sigma \phi$. We have considered the dijets of 2+2, 5+5 and 10+10 particles.
The total particle multiplicity is correspondingly 4, 10, 20 as there is exactly one 
dijet per event and there are no other particles. The results of dijet simulation are 
shown in Fig.~\ref{fig-fifi_jet}.

We have here two sources of azimuthal correlations which counteract each other.
As we already know, the back-to-back emission of particles generates negative
correlations while the collinear emission leads to positive ones. When the particle's 
multiplicity of dijets is sufficiently high and $\sigma \phi$ is sufficiently small, the effect 
of collinear emission  wins  and $\Phi_\phi$ is positive.

%%%%%%%%%%%%%%%%%%%%%%%%%%%%%%%%%%%%%%%%%%%%%%%%%%%%%
\section{Momentum conservation}
%%%%%%%%%%%%%%%%%%%%%%%%%%%%%%%%%%%%%%%%%%%%%%%%%%%%%

The momentum conservation obviously leads to inter-particle correlations.
We have studied the effect on $\Phi_\phi$, generating the sets of particles
of multiplicity $N$. The azimuthal angle distribution of a single particle
is assumed to be flat while the transverse momentum distribution is chosen
in the form
\be
P(p_T) = \beta^2 p_T e^{-\beta p_T }
\ee
with the slope parameter $\beta^{-1}=200$ MeV. For each particle the $x$
and $y$ components of its momentum have been computed as
$p_x = p_T {\rm cos}\phi$ and $p_y = p_T {\rm sin}\phi$. To make the total
transverse momentum of $N$ particles vanish, the $x$ and $y$ component 
of momentum of each particle has been shifted as
\be
p_x \rightarrow p_x - \frac{1}{N}\sum_{i=1}^{N}p_x^i ,
\;\;\;\;\;
p_y \rightarrow p_y - \frac{1}{N}\sum_{i=1}^{N}p_y^i .
\ee

The simulation showing the effect of transverse momentum conservation is
illustrated in Fig.~\ref{fig-zach}. The particle multiplicity has been generated 
according to NB distribution with $\sqrt{{\rm Var}(N)} = \langle N \rangle /2$.
As seen, the effect of momentum conservation is sizable and it survives to
large multiplicities.

In real experiments only a fraction of all produced particles is observed
due to a finite detector efficiency and acceptance. We model the effect of
detector efficiency by randomly loosing particles independently of their
azimuthal angle. In Fig.~\ref{zach_acc} we show how the effect of detector efficiency 
modifies the correlations caused by the transverse momentum conservation. As seen, 
the random losses of particles lead to the dilution of the correlation that is $\Phi_\phi$ 
monotonically goes to zero as the fraction of registered particles 
$f_{\rm reg} \rightarrow 0$.

%%%%%%%%%%%%%%%%%%%%%%%%%%%%%%%%%%%%%%%%%%%%%%%%%%%%%%%%%%%%%%%%
\section{proton-proton collisions in PYTHIA}
%%%%%%%%%%%%%%%%%%%%%%%%%%%%%%%%%%%%%%%%%%%%%%%%%%%%%%%%%%%%%%%%

After the discussion of various mechanisms responsible for azimuthal correlations, 
let us now consider more realistic situation where several mechanisms of azimuthal 
correlations are present at the same time.   We used the PYTHIA generator 
\cite{Sjostrand:2007gs} to simulate p-p collisions at several collision energies 
accessible at SPS ($\sqrt{s_{NN}} = 6.27,\;  7.62,\; 8.73,\;  12.3,\; 17.3$ GeV) 
and RHIC ($\sqrt{s_{NN}} = 19.6,\;  62.4,\; 130,\;  200 $ GeV). For every energy 
a set of minimum bias events was collected. We treated as stable the following particles:  
$\mu^-,\; \pi^0,\; \pi^+,\; K^0,\; K^+,\; K^0_L, \;K^0_S, \;  \Lambda,\; \Sigma^+,\; \Sigma^-,  
\; \Xi^0, \;\Xi^-,\; \Omega^-$ and their antiparticles. No acceptance cuts were applied.

\begin{figure*}[t] 
\vspace{-1.0cm}
\begin{minipage}{8.5cm}
\vspace{-0.3cm}
\center 
\includegraphics*[height=8.5cm]{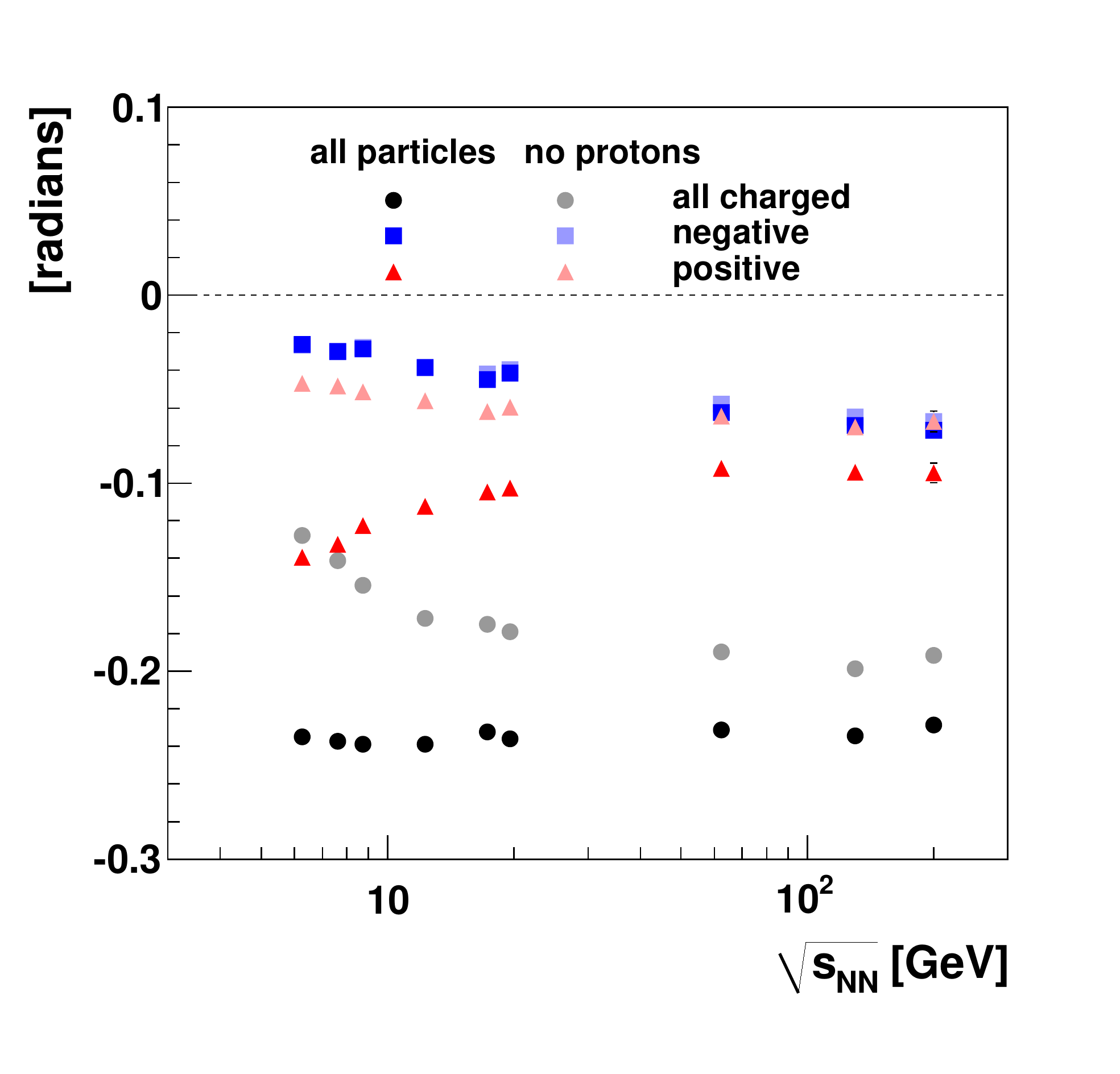}
\vspace{-1.3cm}
\caption{(Color online) The energy dependence of $\Phi_\phi$ for positive, 
negative and all charged particles in the PYTHIA simulations of p-p collisions.  
The pale symbols correspond to the results with protons excluded from 
the positive and all charged particles.}
\label{fig-pythia_fifi}
\end{minipage}\hspace{5mm}
\begin{minipage}{8.5cm}
\center 
\includegraphics*[height=8.5cm]{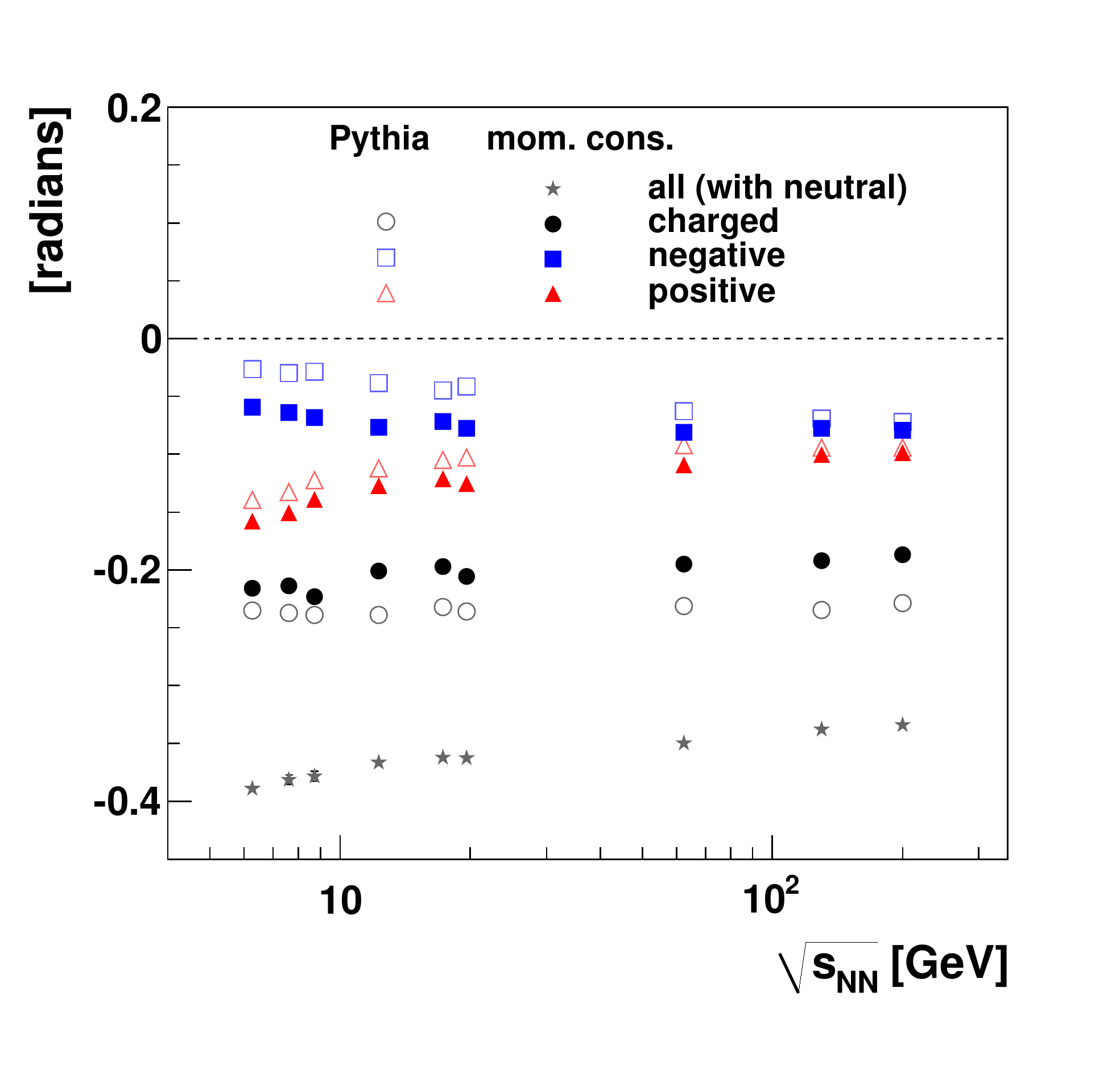}
\vspace{-1.3cm}
\caption {(Color online) $\Phi_\phi$ for the PYTHIA events (open symbols) 
compared to the results of  toy-model simulations (full symbols) which take 
into account only the effect of transverse momentum conservation. The asterisks 
show the toy-model results for all (neutral and charged) particles.}
\label{fig-pT_conserve_fifi}
\end{minipage}
\end{figure*}

For every energy we computed $\Phi_\phi$ separately for positive, negative and all 
charged particles. The results  are shown in Fig.~\ref{fig-pythia_fifi}. As seen, $\Phi_\phi$ 
is negative and weakly depends on collision energy.  To understand why $\Phi_\phi$ is so 
different for negative and for positive particles, we excluded protons from all charged 
and from positive particles. The corresponding values of  $\Phi_\phi$ are also shown in 
Fig.~\ref{fig-pythia_fifi}.  After excluding protons, the correlations among negative particles and among positive are very similar to each other.

What is the mechanism responsible for negative values of $\Phi_\phi$? 
We first checked that high $p_T$ particles play no role here, as $\Phi_\phi$
does not significantly change when particles with $p_T > 1.5$ GeV are
excluded. The correlations among charged particles can be caused by
the effect resonance decays but the effect is certainly very minor for 
same-sign particles, as there are very a few resonances decaying into 
two positive or two negative particles. 

The transverse momentum conservation, which is discussed in Sec.~VI, 
is another possible source of negative values of $\Phi_\phi$. We checked 
that the  PHYTHIA events indeed obey the transverse momentum conservation. 
Specifically, we proved vanishing of the total momentum in $x$ and  in $y$ 
directions of all particles (charged and neutral) from every event. 
To quantitatively study the effect of transverse momentum conservation 
we proceeded as follows. For every collision energy we determined the
average multiplicity of positive, negative and neutral particles. Then, we
performed the simple simulations described in Sec.~VI, generating events 
of a given total multiplicity which satisfy the transverse momentum conservation.
Then, a fraction of particles was randomly eliminated to get the multiplicity
of charged, positive or negative particles.  The values of $\Phi_\phi$ 
computed for such events are shown in Fig.~\ref{fig-pT_conserve_fifi}.
As seen,  the values of $\Phi_\phi$ for the PYTHIA events agree quite well 
with the results of our toy-model simulations which take into account only 
the effect of transverse momentum conservation. It is somewhat surprising 
that the agreement for same-sign particles is not much better that that for all 
charged particles. It means that the resonance decays do not generate strong 
correlations in the PYTHIA events. We note, however, that the effect of transverse
momentum conservation overshoots the correlations of the like-sign particles
and it undershoots the correlations of all charged particles. The latter results 
presumably signals presence of resonances decaying into pairs of one positive 
and one negative particle. 

%%%%%%%%%%%%%%%%%%%%%%%%%%%%%%%%%%%%%%%%%%%%%%%%%%%%%%%%%%%%%%%%
\section{nucleus-nucleus collisions in  HIJING}
%%%%%%%%%%%%%%%%%%%%%%%%%%%%%%%%%%%%%%%%%%%%%%%%%%%%%%%%%%%%%%%%

\begin{figure}[b] 
\centering
\vspace{-0.5cm}
\includegraphics*[width=0.49\textwidth]{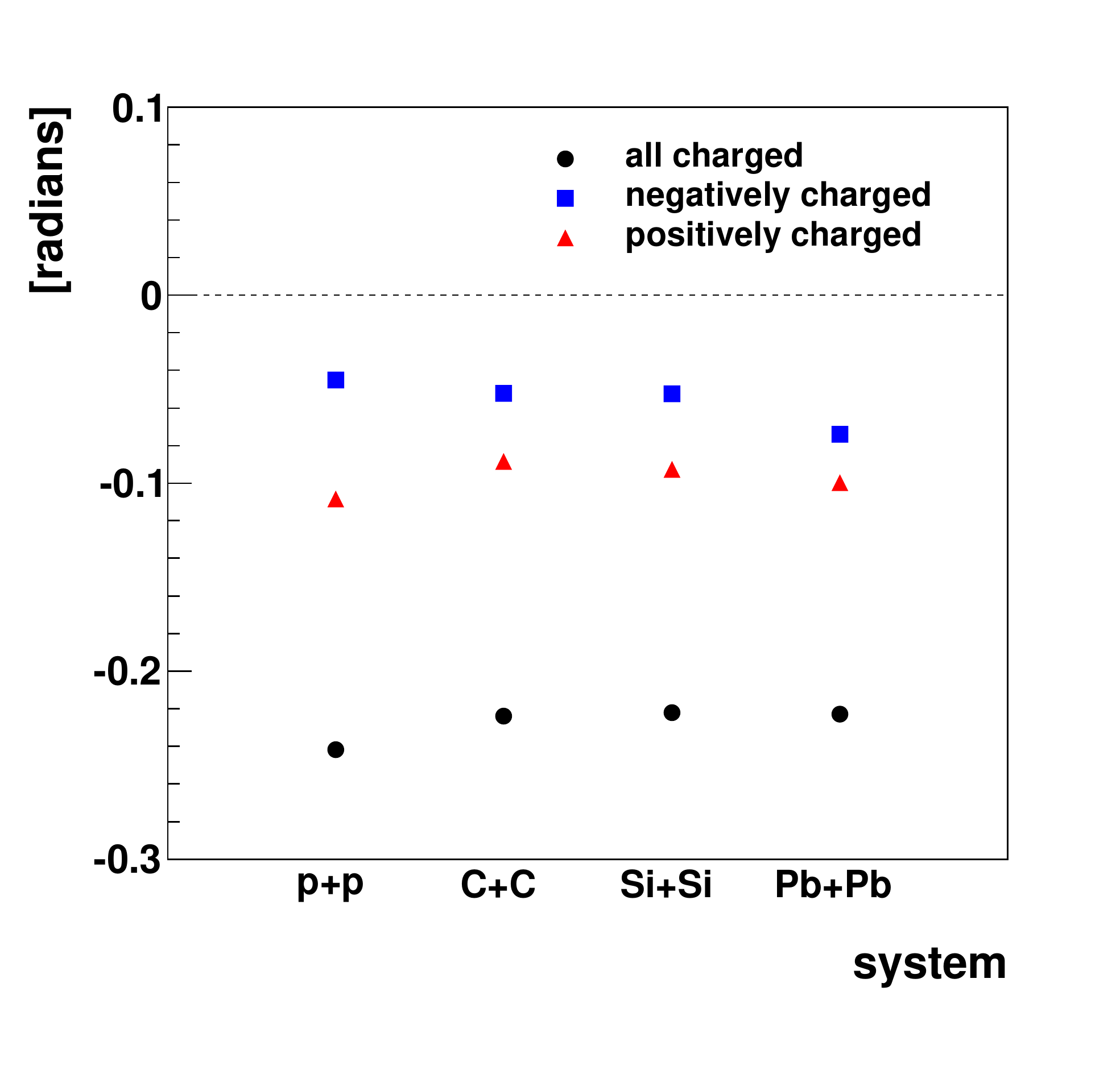}
\vspace{-1.4cm}
\caption {The Hijing simulation of the system size dependence 
of $\Phi_\phi$ for nucleus-nucleus collisions at 
$\sqrt{s_{NN}} = 17.3$ GeV with no acceptance cuts.}
\label{fig-hijinig_fifi_system}
\end{figure}

We have also performed simulations of nucleus-nucleus collisions
using the HIJING \cite{Gyulassy:1994ew} event generator.  We have simulated 
the collisions of p-p, C-C, Si-Si and Pb-Pb at $\sqrt{s_{NN}} = 17.3$ GeV and 
$\Phi_\phi$ has been computed separately for positive, negative and all charged 
particles coming from minimum bias events. The results are shown in 
Fig.~\ref{fig-hijinig_fifi_system}. As seen, $\Phi_\phi$ is almost independent 
of the mass number of colliding nuclei and the values of $\Phi_\phi$ are very 
close to those found using PYTHIA. It is by no means accidental.  When 
a nucleus-nucleus collision is a simple superposition of nucleon-nucleon
interactions, the value of $\Phi$ is exactly the same for p-p interactions and 
nucleus-nucleus collisions at any centrality.  In the HIJING model a nucleus-nucleus 
collision is not exactly a superposition of nucleon-nucleon collisions but it is almost so. 
And the treatment  of proton-proton interactions is essentially the same in PYTHIA and 
HIJING. For these reasons our analysis of PYTHIA events presented in the previous
section fully applies here.

%%%%%%%%%%%%%%%%%%%%%%%%%%%%%%%%%%%%%%%%%%%%%%%%%%%%%%%%%%%%%%%%
\section{Summary and outlook}
%%%%%%%%%%%%%%%%%%%%%%%%%%%%%%%%%%%%%%%%%%%%%%%%%%%%%%%%%%%%%%%%

Azimuthal correlations of final state particles from high-energy collisions carry valuable 
information on the collision dynamics. It motivates the analysis of experimental data 
collected by the NA49 and NA61 Collaborations which is in progress with some preliminary 
results already published \cite{Cetner:2010vz}. The integral measure $\Phi_\phi$, which 
proved to be very sensitive to various dynamical correlations, is used in the analysis. 
To interpret the experimental results it should be understood how different sources of 
correlations manifest themselves when measured by means of $\Phi_\phi$. This was 
the aim of our study. We performed several simulations to analyze separately 
the azimuthal correlations  caused by the elliptic flow, resonance decays, jets and 
transverse momentum conservation. We also discussed how the correlations are 
diluted due to randomly lost particles. Finally we used the PYTHIA and HIJING 
event generators to produce a big sample of events which mimic experimental data 
from p-p and nucleus-nucleus collisions at the SPS and RHIC collision energies.  
$\Phi_\phi$ appeared to be surprisingly independent of  the collision energy and of
the size of colliding systems. Applying some kinematic cuts and selection criteria of particles, 
we showed that the azimuthal correlations are dominated by rather trivial effect 
of transverse momentum conservation which appeared to be almost independent 
of particle's multiplicity which changes dramatically for collision energies and system's 
sizes under consideration. 

The experience gathered  in the course of this theoretical study  will be used to better 
understand experimental data. Quantitative analysis of several simple mechanisms of 
azimuthal correlations we discussed will facilitate an observation of possible new 
phenomena like critical fluctuations at phase boundaries of strongly interacting matter 
or plasma instabilities from the early stage of relativistic heavy-ion collisions. 

%-----------------------------------------------------------------------
\section*{Acknowledgments}
%-----------------------------------------------------------------------

We are grateful to Maciej Rybczy\'nski for providing us with the sample of nucleus-nucleus 
collisions at $\sqrt{s_{NN}} = 17.3$ GeV simulated with the HIJING event generator. 
We thank to Marek Ga\' zdzicki for critical reading of the manuscript and his numerous
comments. This work was partially supported by Polish Ministry of Science  and Higher  
Education under grants N~N202~204638 and 667/N-CERN/2010/0.

%-----------------------------------------------------------------------
\appendix*
%-----------------------------------------------------------------------
%%%%%%%%%%%%%%%%%%%%%%%%%%%%%%%%%%%%%%%%%%%%%%%%%%%%%%%%%%%%%%%%%%%

\section{Toy model of resonance decays}

%%%%%%%%%%%%%%%%%%%%%%%%%%%%%%%%%%%%%%%%%%%%%%%%%%%%%%%%%%%%%%%%%%%

The inclusive distribution of azimuthal angle is assumed to be flat that is
\be
\label{incl}
P_{\rm inc}(\phi) = 
{1 \over 2\pi} \; \Theta(\phi) \, \Theta(2\pi - \phi) \;,
\ee
which gives $\bar \phi = \pi$ and $\bar \phi^2 = 4\pi^2/3$.
Consequently, $\bar z^2 = \pi^2/3$.

Let us first assume that {\em all} produced particles come from heavy 
resonances which are at rest and decay back to back into two particles. 
When one particle is emitted at the azimuthal angle $\phi_1$ and 
$0 \le \phi_1 < \pi$, the second particle is emitted at 
$\phi_2 = \phi_1 + \pi$. When $\pi \le \phi_1 < 2\pi $, then
$\phi_2 = \phi_1 - \pi$. Therefore, the two-particle distribution 
of azimuthal angles reads
\ba
\label{2-particle-decays}
P_2(\phi_1, \phi_2) &=& 
{1 \over 2\pi} \; 
\Theta(\pi -\phi_1) \, \delta(\phi_1 - \phi_2 +\pi)
\\ \nonumber
&+&
{1 \over 2\pi} \; 
\Theta(\phi_1 - \pi) \, \delta(\phi_1 - \phi_2 -\pi)
\;.
\ea
One observes that
\be
\int d\phi_1 P_2(\phi_1, \phi) 
= \int d\phi_2 P_2(\phi, \phi_2) = P_{\rm inc}(\phi) \;,
\ee
and computes 
\be
\label{ave-phi-1-phi-2}
\int d\phi_1 d\phi_2 \phi_1 \phi_2 P_2(\phi_1, \phi_2) 
= \frac{5}{6} \pi^2
\;.
\ee 

\begin{widetext}

We further assume that the particle multiplicity is arbitrary
but fixed even number $N$. Then, the $N-$particle distribution
of azimuthal angles is
\be
\label{N-particle-decays}
P_N(\phi_1, \phi_2, \dots , \phi_N ) = 
P_2(\phi_1, \phi_2) \:
P_2(\phi_3, \phi_4) \: \cdots \:
P_2(\phi_{N-1}, \phi_N ) \;.
\ee
The variable $Z$ is defined as $Z = \phi_1+\phi_2 + \dots +\phi_N -N\pi$
and one computes $\langle Z^2 \rangle$ in the following way
\ba
\label{Z^2-1}
\langle Z^2 \rangle
&=& \int d\phi_1 d\phi_2\dots d\phi_N
(\phi_1+\phi_2 + \dots +\phi_N -N\pi)^2 
P_N(\phi_1, \phi_2, \dots , \phi_N )
\\ [2mm] \nonumber
&=& N \bar{\phi^2} 
+ N \int d\phi_1 d\phi_2 \phi_1 \phi_2 P_2(\phi_1, \phi_2)
+ N(N-2) {\bar\phi}^{2} 
-2N^2 \bar\phi \pi + N^2 \pi^2
= \frac{N}{6} \pi^2 \;.
\ea
Using the result (\ref{Z^2-1}) and keeping in mind that $\bar z^2 = \pi^2/3$, 
we find the formula (\ref{Phi-resonance-decays}) from the definition 
(\ref{phi-def}).

Let us now assume that $N_1$ particles come from  resonances and additional
$N_2$ particles are produced independently from each other and from 
resonances. Then, we still have $\bar z^2 = \pi^2/3$ and 
$\langle Z^2 \rangle$ is computed as
\ba
\label{Z^2-2}
\langle Z^2 \rangle
&=& N \bar{\phi^2} 
+ N_1 \int d\phi_1 d\phi_2 \phi_1 \phi_2 P_2(\phi_1, \phi_2)
+ N_1(N_1-2) {\bar\phi}^2 
\\ [2mm] \nonumber
&+& 2 N_1 N_2  {\bar\phi}^2 
+N_2(N_2-1) {\bar\phi}^2
-2N^2 \bar\phi \pi + N^2 \pi^2
= \frac{\pi^2}{6}\,(2N -N_1) \;,
\ea
where $N \equiv N_1 +N_2$. The result (\ref{Z^2-2}) with $f\equiv N_1/N$
gives the formula (\ref{Phi-resonance-decays-fraction}).

The model can be easily generalized to the situation when the particles from a correlated
pair are not emitted back to back but their relative azimuthal angle is fixed and equal 
$\Delta \phi$. Then,  the two-particle distribution of azimuthal angles is
\ba
\label{2-particle-decays-general}
P_2(\phi_1, \phi_2) 
= {1 \over 2\pi} \; 
\Theta(2\pi - \Delta \phi - \phi_1) \, \delta(\phi_1 - \phi_2 + \Delta \phi)
+ {1 \over 2\pi} \; 
\Theta(\phi_1 - 2\pi + \Delta \phi) \, \delta(\phi_1 - \phi_2 + \Delta \phi - 2\pi)
\;,
\ea

\end{widetext}

and instead of Eq.~(\ref{ave-phi-1-phi-2}) we have
\be
\label{ave-phi-1-phi-2-general}
\int d\phi_1 d\phi_2 \phi_1 \phi_2 P_2(\phi_1, \phi_2) 
= \frac{4}{3} \pi^2 - \Delta \phi \, \pi + \frac{1}{2} (\Delta \phi)^2 \;.
\ee 
Substituting the result (\ref{ave-phi-1-phi-2-general}) to Eq.~(\ref{Z^2-1}) or 
Eq.~(\ref{Z^2-2}), one finds $\langle Z^2 \rangle$ which finally gives 
the formula (\ref{Phi-resonance-decays-general}) or 
(\ref{Phi-resonance-decays-general-f}), respectively.

\end{document}